\documentclass[apj]{emulateapj}

\usepackage{natbib} 
\usepackage{graphicx}
\usepackage{lipsum}
\usepackage{threeparttable}
\usepackage{amsmath}
\usepackage{color}
\usepackage{booktabs}
\usepackage{multirow} 

\definecolor{orange}{rgb}{0.93, 0.57, 0.13}
\definecolor{green}{rgb}{0.01, 0.75, 0.24}

\newcommand{\ie}{\emph{i.e.}}

\newcommand{\emm}[1]{\ensuremath{#1}}   
\newcommand{\emr}[1]{\emm{\mathrm{#1}}} 

\newcommand{\Msun}{\emm{M_\odot}}
\newcommand{\Rout}{\emm{R_\emr{out}}}

\newcommand{\hthcn}{\emr{H^{13}CN}}                  
\newcommand{\hcfin}{\emr{HC^{15}N}}                  

\newcommand{\Nratio}{\emr{^{14}N/^{15}N}}                   %

\renewcommand{\deg}{\emm{^\circ}}

\newcommand{\pscm}{~\rm{cm}^{-2}}

\newcommand{\kms}{\emr{\,km\,s^{-1}}}

\shorttitle{Cyanide photochemistry and fractionation in the MWC 480 disk}
\shortauthors{V.V. Guzm\'an et al.}

\begin{document}


\title{Cyanide photochemistry and Nitrogen fractionation in the MWC 480 disk}


\author{V.V. Guzm\'an\altaffilmark{1}}
\email{vguzman@cfa.harvard.edu}

\author{K.I. \"Oberg\altaffilmark{1}}

\author{R. Loomis\altaffilmark{1}}

\author{C. Qi\altaffilmark{1}}

\altaffiltext{1}{Harvard-Smithsonian Center for Astrophysics, 60 Garden Street, Cambridge, MA 02138, USA}

\begin{abstract}
HCN is a commonly observed molecule in Solar System bodies and in
interstellar environments. Its abundance with respect to CN is a
proposed tracer of UV exposure. HCN is also frequently used to probe
the thermal history of objects, by measuring its degree of nitrogen
fractionation. To address the utility of HCN as a probe of disks, we
present ALMA observations of CN, HCN, H$^{13}$CN and HC$^{15}$N toward
the protoplanetary disk around Herbig Ae star MWC480, and of CN and
HCN toward the disk around T Tauri star DM Tau. Emission from all
molecules is clearly detected and spatially resolved, including the
first detection of HC$^{15}$N in a disk. Toward MWC 480, CN emission
extends radially more than 1$"$ exterior to the observed cut-off of
HCN emission. Quantitative modeling further reveals very different
radial abundance profiles for CN and HCN, with best-fit outer cut-off
radii of $>$300~AU and 110$\pm$10~AU, respectively. This result is in
agreement with model predictions of efficient HCN photodissociation
into CN in the outer-part of the disk where the vertical gas and dust
column densities are low. No such difference in CN and HCN emission
profiles are observed toward DM Tau, suggestive of different
photochemical structures in Herbig Ae and T Tauri disks.  We use the
HCN isotopologue data toward the MWC 480 disk to provide the first
measurement of the \Nratio{} ratio in a disk. We find a low disk
averaged \Nratio{} ratio of 200$\pm$100, comparable to what is
observed in cloud cores and comets, demonstrating interstellar
inheritance and/or efficient nitrogen fractionation in this disk.
\end{abstract}


 \keywords{astrochemistry, ISM: molecules, protoplanetary disks, radio lines; techniques: high angular resolution}


\newcommand{\TabObs}{%
  \begin{table*}
    \begin{center}
    {\small \caption{Observation parameters.}}
    \label{tab:obs}
        \begin{tabular}{lcccccc}\toprule
          Line & Frequency & $E_u/k$ & Beam & PA & Channel width & Noise \\
          & GHz & K & arcsec & $^{\deg}$ & \kms{} & mJy beam$^{-1}$ channel$^{-1}$\\ 
          \midrule
          CN  $N=3-2, J=7/2-5/2$ & 340.24777 & 32.6 & $0.78\times0.45/0.65\times0.48$ & $-23.1/-29.2$ & 0.1 & $12.3/9.2$\\
          HCN $J=4-3$            & 354.50548 & 42.5 & $0.77\times0.41/0.64\times0.40$ & $-24.1/-29.0$ & 0.1 & $14.0/11.2$\\
          \hthcn{} $J=3-2$         & 259.0118 & 24.9 & $0.74\times0.46$ & $-8.8''$ & 0.2 & $3.2$\\
          \hcfin{} $J=3-2$         & 258.1571 & 24.8 & $0.74\times0.46$ & $-9.0''$ & 0.2 & $4.2$ \\
          \bottomrule
        \end{tabular}
    \end{center}
    Note: Two values are given for the beam, PA and noise, corresponding to the
    parameters in MWC~480 and DM~Tau, respectively.
\end{table*}
}

\newcommand{\TabParam}{%
  \begin{table}[t!]
    {\small \caption{Adopted parameters in disk model for MWC~480.}}
    \label{tab:param}
        \begin{tabular}{llll}\toprule
          \multicolumn{2}{c}{Parameter} & Value & Ref. \\
          \midrule
          \multirow{2}{*}{Scale height} & $H_{100}$ & 16~AU & \multirow{2}{*}{\cite{guilloteau2011}}\\
          & $h$ & 1.25\\
          \midrule
          \multirow{3}{*}{Surface density}  & $M_{gas}$ & 0.18\Msun{} & \multirow{3}{*}{\cite{guilloteau2011}} \\
           & $\gamma$ & 0.75\\
          & $R_c$ & 81~AU\\
          \midrule
          \multirow{3}{*}{Temperature} & $T_{100}$ & 23~K & \multirow{2}{*}{\cite{pietu2007}}\\
          & $q$ & 0.5\\
          & $\beta$ & 1.5 & \cite{dartois2003}\\
          \bottomrule
        \end{tabular}
\end{table}
}

\newcommand{\TabModelResults}{%
  \begin{table*}[t]
    \begin{center}
      {\small \caption{Best-fit parameters for the modeled abundance
          profiles in MWC~480.} }
      \label{tab:modelresults}
      \begin{tabular}{ccccc}\toprule
        & CN & HCN & \hthcn{} & \hcfin{} \\
        \midrule
        $X_0$   & $(8.5\pm1.5)\times10^{-11}$ & $(5.2\pm0.7)\times10^{-12}$  & $(1.3\pm0.5)\times10^{-13}$ & $(4.7\pm1.7)\times10^{-14}$\\
        $\alpha$ & -1 & -1 & -0.5 & -0.5 \\
        \Rout{} [AU] & $>300$ & $110\pm10$ & $100\pm15$ & $95\pm30$ \\
        \bottomrule
      \end{tabular}
  \end{center}
    Note: The absolute abundances are largely uncertain, due to
    the unknown vertical abundance and temperature profiles. However,
    because \hthcn{} and \hcfin{} are expected to be, for the most
    part, co-spatial their abundance ratio does not suffer
    greatly from uncertainties in the vertical temperature
    profile. Therefore, the inferred $\hthcn/\hcfin$ ratios is
    robust.
  \end{table*}
}

\newcommand{\FigChannelMaps}{%
\begin{figure*}[t!]
  \centering
  \includegraphics[width=\textwidth]{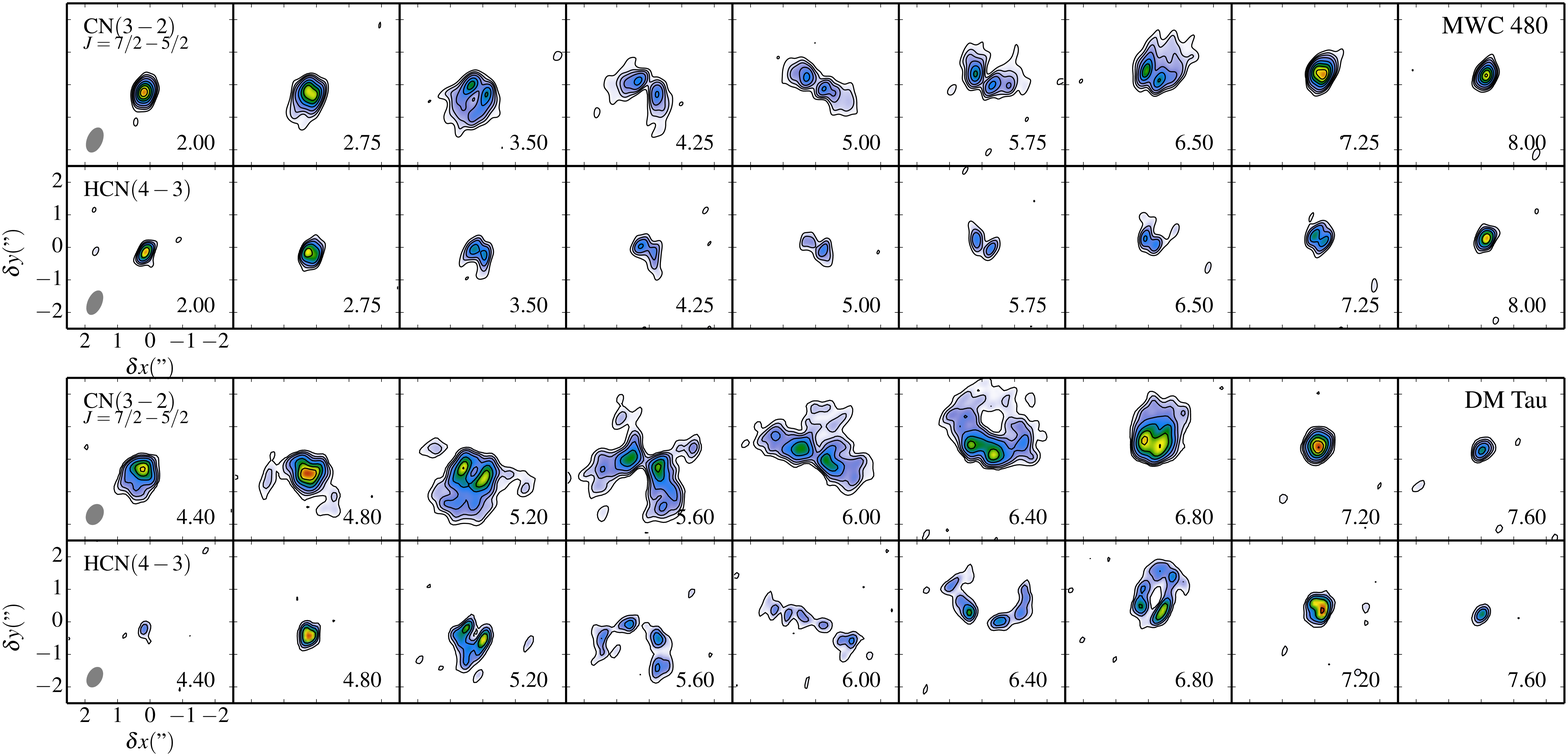}
  \caption{Channel maps of the HCN(4-3) and CN(3-2) line emission in
    MWC~480 (top) and DM~Tau (bottom). The contour levels correspond
    to 3, 5, 7, 10, 15, 20 and 25$\sigma$ (Table
    \ref{tab:obs}). Synthesized beams are shown in the bottom left
    corner of the first panels.}
  \label{fig:channelmaps}
\end{figure*}
}

\newcommand{\FigRadProfiles}{%
\begin{figure*}[t!]
  \centering
  \includegraphics[width=0.5\textwidth]{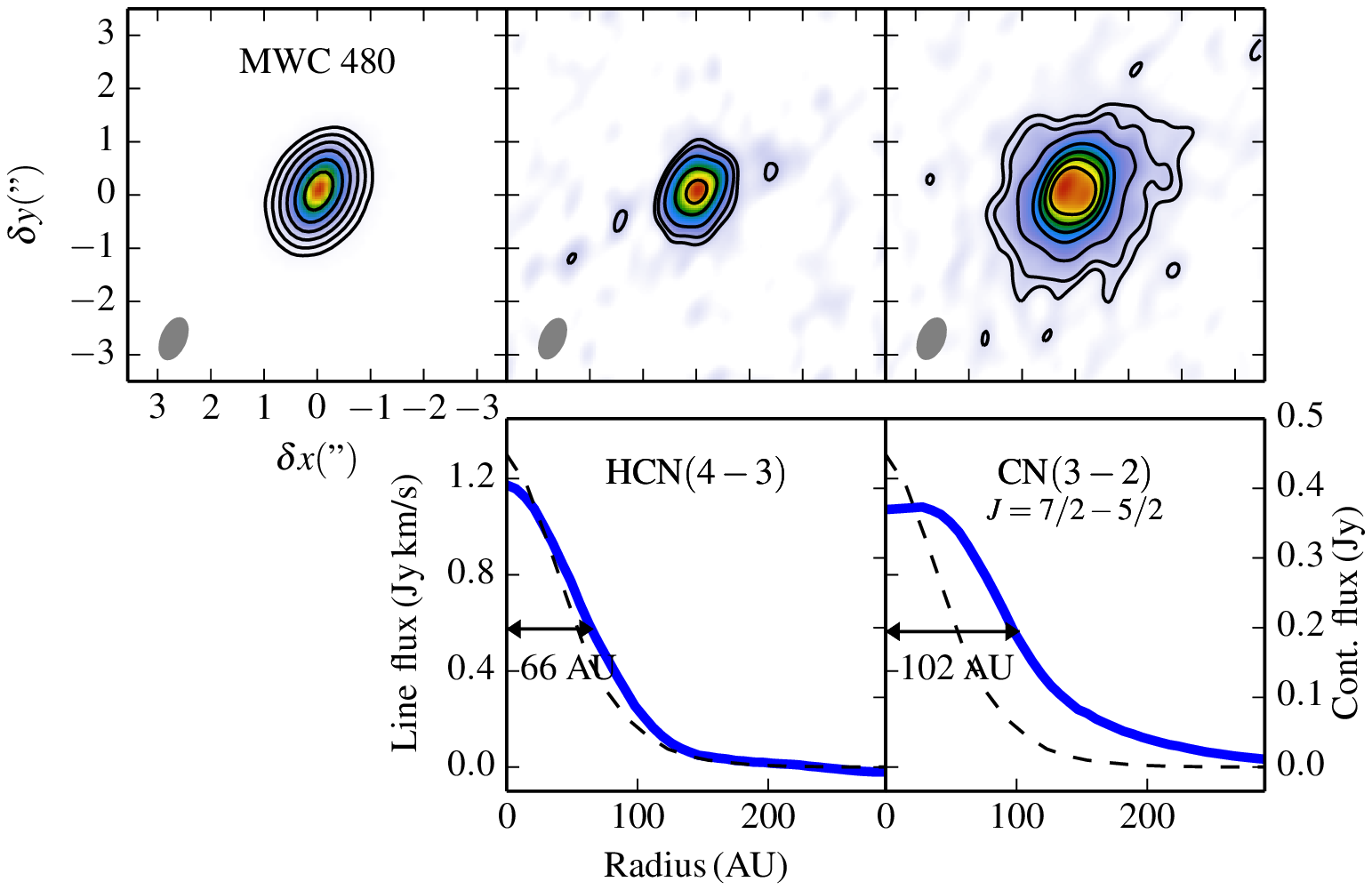}\hspace{-0.3cm}
  \includegraphics[width=0.5\textwidth]{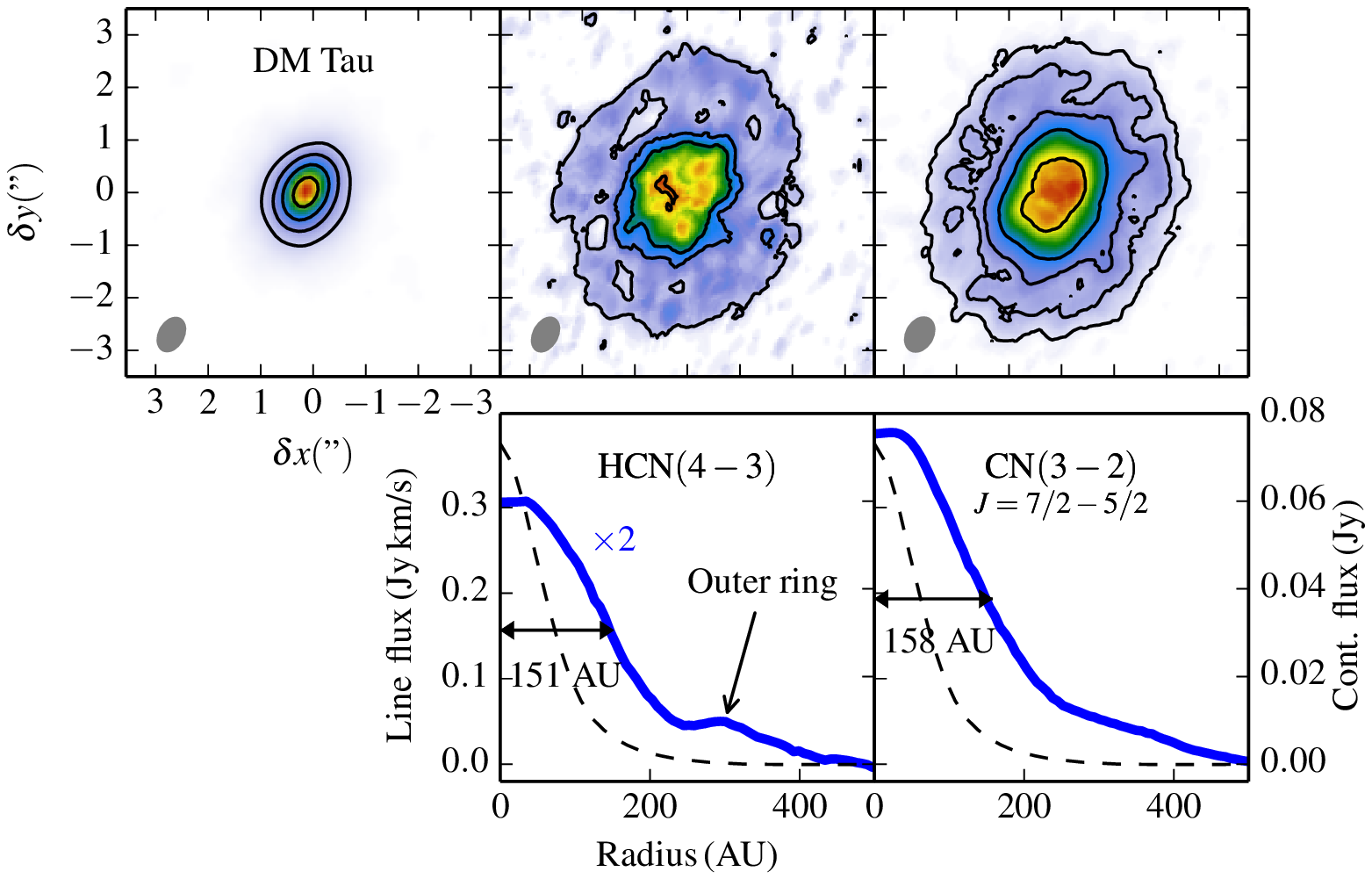}
  \caption{Dust continuum emission, moment zero maps (upper panels)
    and azimuthally-averaged emission profiles (bottom panels) of the
    HCN $J=4-3$ and CN $N=3-2$, $J=7/2-5/2$ lines (blue solid lines)
    in MWC480 (left) and DM Tau (right). The azimuthally-averaged
      dust continuum profiles are shown in dashed lines. The weaker
    HCN line in DM~Tau has been multiplied by 2 to show the radial
    profile, and the outer ring near 300~AU, more clearly. Synthesized
    beams are shown in the bottom left panel corners. The
    double-arrows mark the half-light radii.}
  \label{fig:profiles}
\end{figure*}
}

\newcommand{\FigResiduals}{%
\begin{figure*}[t!]
  \centering
  \includegraphics[width=\textwidth]{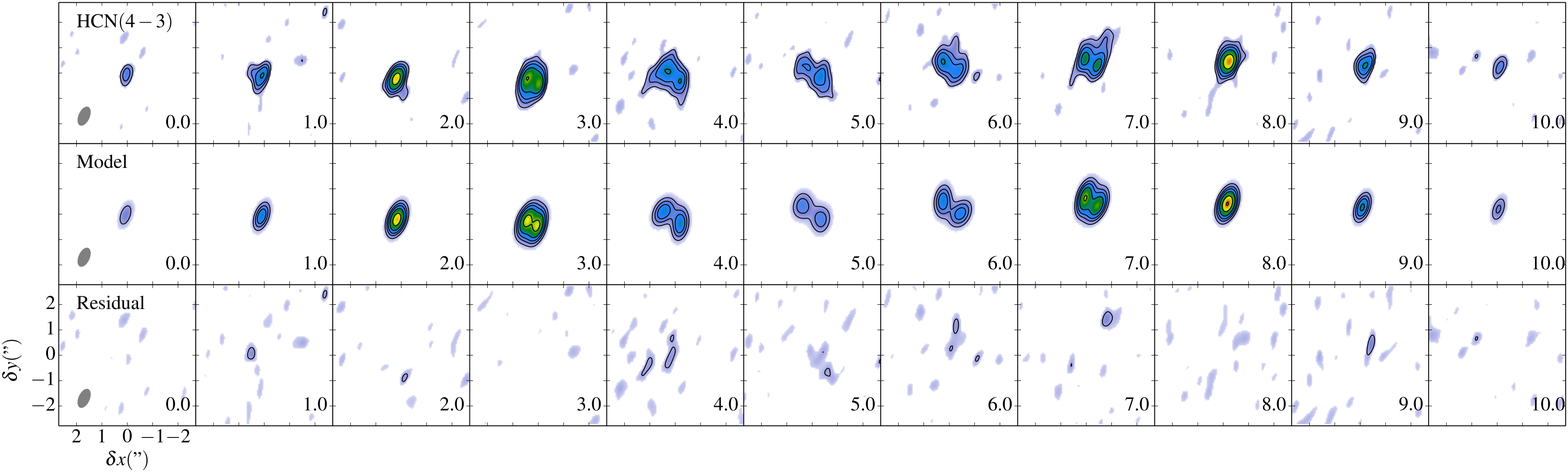}
  \includegraphics[width=\textwidth]{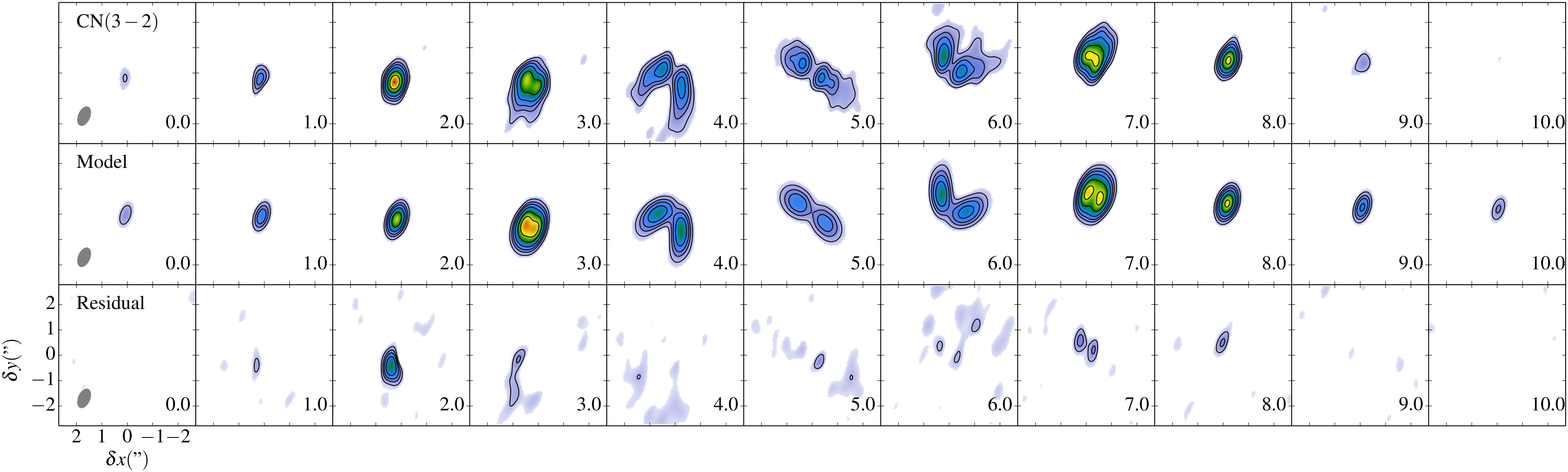}
  \caption{Channel maps of the observed CN and HCN emission in MWC~480
    (upper panels), together with synthetic observations generated
    from the best-fit models for each line, corresponding to
    $X_0=1.0\times10^{-10}$, $\Rout=350$~AU and $\alpha=-1$ for CN,
    and $X_0=5.0\times10^{-12}$, $\Rout=110$~AU and $\alpha=-1$ for
    HCN (middle panel). The residuals are shown in the bottom
    panels. The contour levels correspond to 3, 5, 7, 10, 15, 20 and
    25$\sigma$.}
  \label{fig:residuals}
\end{figure*}
}

\newcommand{\FigIsotpologues}{%
\begin{figure}[t!]
  \centering
  \includegraphics[width=0.43\textwidth]{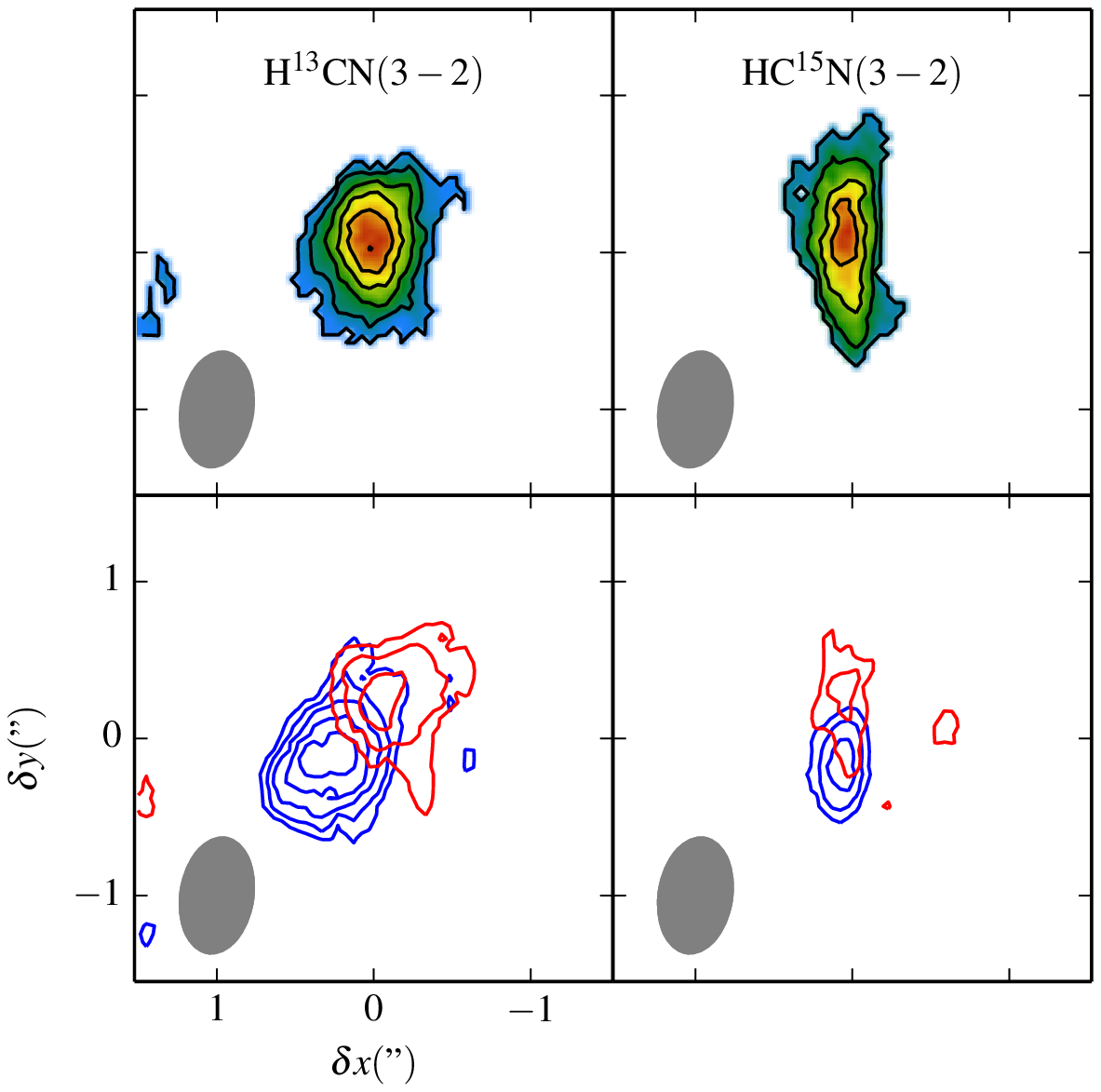}
  \caption{Moment zero maps (left panels) of the $\hthcn$ $J=3-2$ and
    $\hcfin$ $J=3-2$ lines in the MWC~480 disk, shown together with integrated
    flux maps in two velocity bins around the $v_{lsr}$ (right
    panels), showing the velocity pattern.}
  \label{fig:isotopologues}
\end{figure}
}

\newcommand{\FigIsotopologuesRatios}{%
\begin{figure}[t!]
  \centering
  \includegraphics[width=0.45\textwidth]{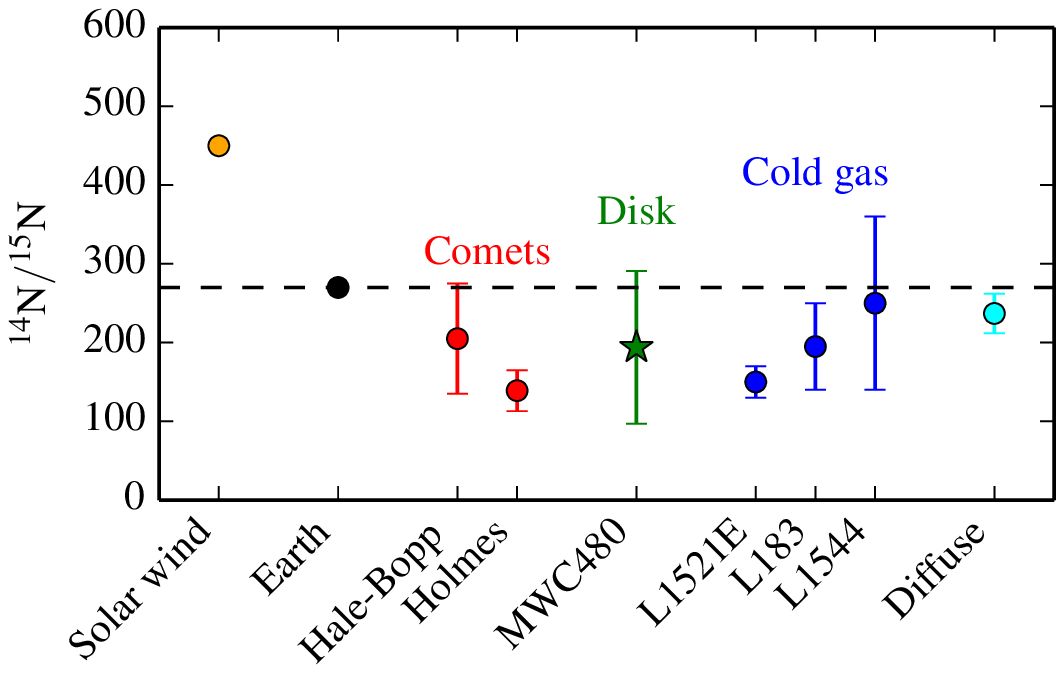}
  \caption{\Nratio{} isotopic ratios observed towards different Solar
    system bodies \citep{marty2011,bockelee2008} and
    in the ISM \citep{hily-blant2013,lucas1998}, and the disk around MWC~480.}
  \label{fig:ratios}
\end{figure}
}

\newcommand{\FigChiCN}{%
\begin{figure}[t!]
  \centering
  \includegraphics[width=0.47\textwidth]{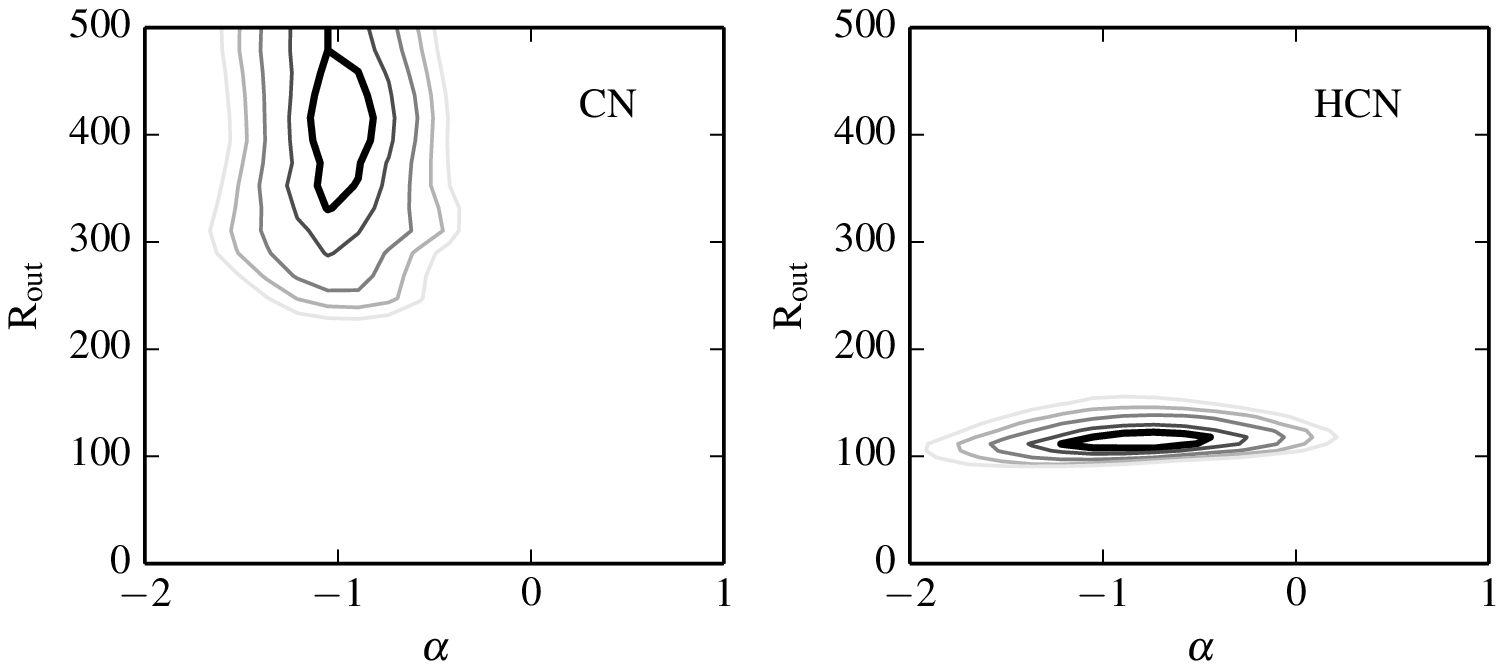}
  \caption{Contour plots of the $\chi^2$ surface for CN (left) and HCN
    (right) abundance profiles in the MWC 480 disk.  The abundance
      at 100~AU ($X_0$) is fixed to the best-fit model value (see
      Table~3). The thick contour corresponds to the $7\sigma$
    contour level, and the light-gray lines mark contours with $\Delta
    \chi^2=100, 200, 300$ and 400.}
  \label{fig:chi_cn_hcn}
\end{figure}
}

\newcommand{\FigChiISotop}{%
\begin{figure}[t!]
  \centering
  \includegraphics[width=0.47\textwidth]{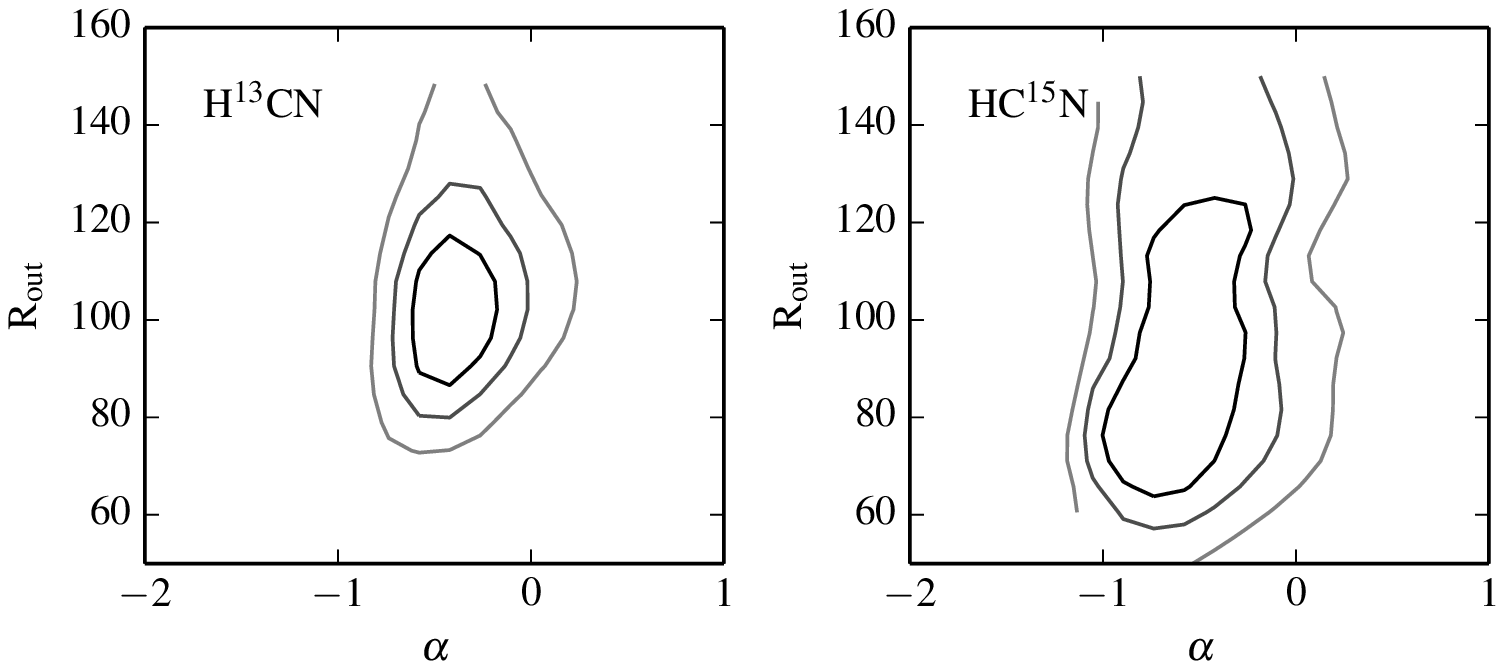}
  \caption{Contour plots of the $\chi^2$ surface for \hthcn{} (left)
    and \hcfin{} (right) abundance profiles in the MWC 480 disk. 
      The abundance at 100~AU ($X_0$) is fixed to the best-fit model
      value (see Table~3). The three
    contours correspond to $1\sigma$, $2\sigma$ and $3\sigma$
    confidence levels.}
  \label{fig:chi_h13cn_hc15n}
\end{figure}
}

\newcommand{\FigSpectraHCN}{%
\begin{figure}[t!]
  \centering
  \includegraphics[width=0.45\textwidth]{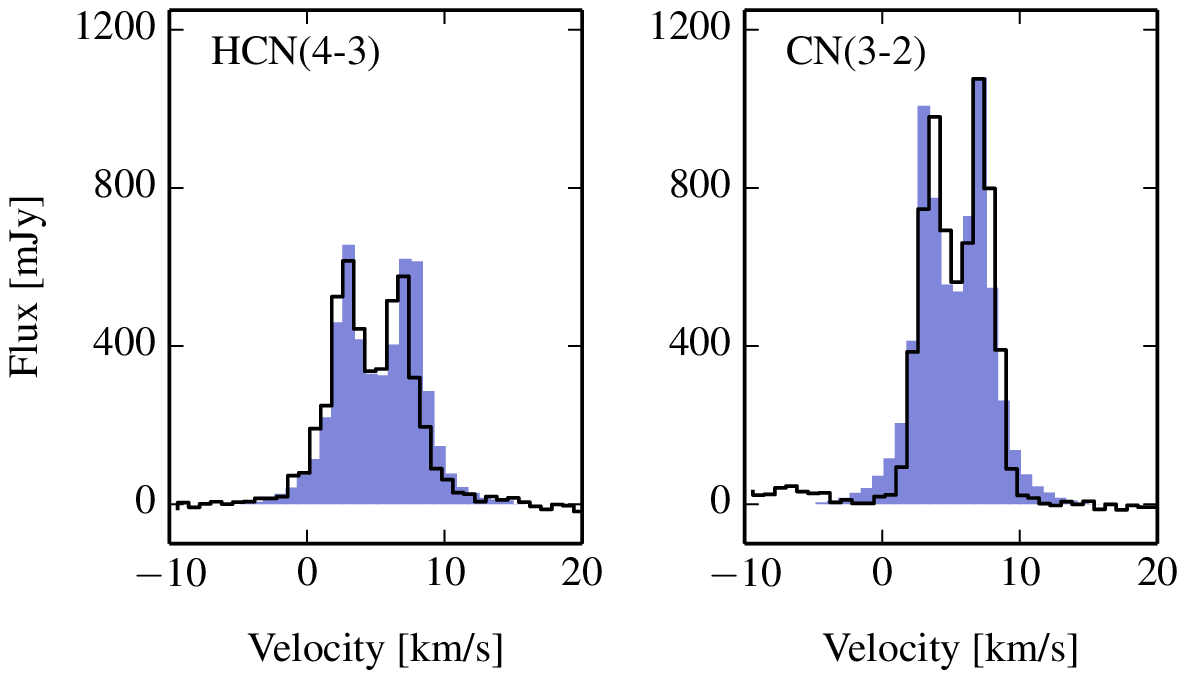} 
  \caption{HCN and CN spectra integrated over the MWC 480 disk (black
    histograms), with the best-fit models overlaid on top (filled
    histograms).}
  \label{fig:spectra_HCN}
\end{figure}
}

\newcommand{\FigSpectraIso}{%
\begin{figure}[t!]
  \centering
  \includegraphics[width=0.45\textwidth]{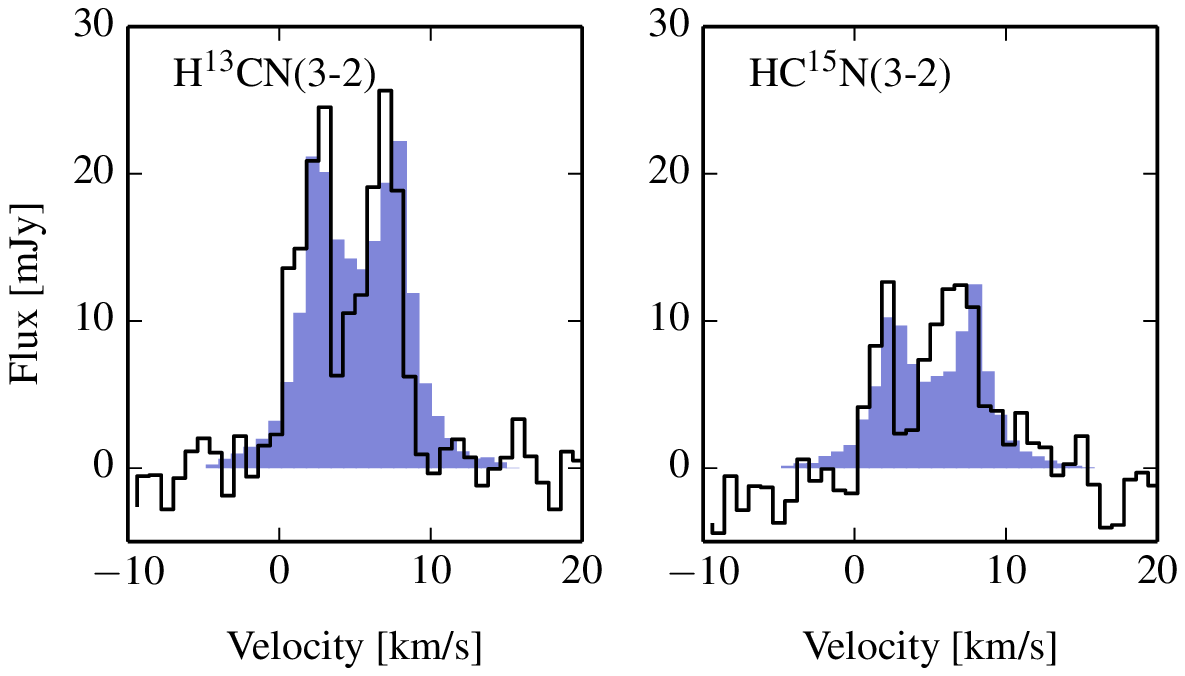}
  \caption{\hthcn{} and \hcfin{} integrated spectra over the MWC 480
    disk (black histograms) with the best-fit models overlaid on top
    (filled histograms).}
  \label{fig:spectra_Iso}
\end{figure}
}


\section{Introduction}
\TabObs{}

Protoplanetary disks are characterized by radial and vertical
temperature and UV radiation gradients due to the exposure of disk
surfaces to stellar and interstellar radiation, and attenuation of
this radiation by dust and gas \citep{calvet1991,herczeg2002}. This
radiation and temperature structure should result in a chemically
stratified disk structure. Atoms and radicals (e.g. CN) alone are
expected to survive in the UV-illuminated disk surface
\citep{bergin2003}. Molecules that are readily photo dissociated by UV
radiation are mainly present in a protected layer between the disk
surface and the cold midplane. In the midplane freeze-out of heavy
atoms is expected to be nearly complete \citep{dutrey1997,qi2013}. The
existence of these cold disk regions may be conductive to efficient
fractionation of heavy isotopes into specific molecules as is
introduced below.

Due to their different levels of photo resistance, CN/HCN ratio has
been proposed as a tracer of UV radiation fields in disks
\citep{fuente1993}. In addition to different vertical scale heights,
\citet{jonkheid2007} proposed that CN emission should be much more
extended than HCN emission in disks around Herbig Ae stars due to UV
penetration down to the midplane in the outer low-density disk. In the
outer disk region, the disk molecular layer is effectively absent,
resulting in a lack of HCN emission. Whether a similar structure
should be expected around T Tauri stars is unclear because Herbig Ae stars
are orders of magnitude more luminous than T Tauri
stars. \citet{oberg2011} used 3\arcsec{} Submillimeter Array (SMA)
observations of CN and HCN in a sample of disks to search for
differences in the CN and HCN emission radii, and identified
low-significance $1-2\sigma$ differences in a handful of
cases. \citet{chapillon2012} observed CN and HCN emission at an
angular-resolution of $1''.5-3''$ towards two T Tauri stars and one
Herbig Ae star, but could not constrain the relative radial
distribution because of different excitation characteristics of the
observed HCN and CN lines.

HCN isotopologue ratios are potential probes of the disk thermal
history. HCN is expected to become enriched in $^{15}$N (chemically
fractionated) at low temperatures due to a small difference in
zero-point energies for the $^{14}$N and $^{15}$N isotopologues
\citep{terzieva2000,rodgers2008}. Based on this scenario, the nitrogen
isotopic compositions of different bodies in the Solar System (the
Sun, rocky planets, gaseous planets, comets, meteorites, etc) have
been used to understand their origin \citep[see ][and references
  therein]{mumma2011}. There are, however, other potential sources of
$^{15}$N enrichment than chemical fractionation \citep[][and see
  Discussion]{heays2014}. Measurements of nitrogen fractionation in
analogs to the Solar Nebula are therefore key to interpreting Solar
System isotopic data.

\FigChannelMaps{}

In this paper we present ALMA observations of CN and HCN, and the HCN
isotopologues H$^{13}$CN and HC$^{15}$N towards the protoplanetary
disks around $1.8 \Msun$ Herbig Ae star MWC~480. We also present CN
and HCN data toward the disk around $0.5 \Msun$ T Tauri star DM~Tau to
enable a comparison between T Tauri and Herbig Ae photochemistry. Both
stars are located in the Taurus star forming region at a distance of
140~pc, and are known to present large, chemically rich disks
\citep{dutrey1997,henning2010,oberg2010,chapillon2012b,oberg2015}.
The observations are data reduction process are described in
section~\ref{sec:obs}. The observational results and analysis are
presented in section~\ref{sec:results} and a discussion is given in
section~\ref{sec:discussion}. We summarize our findings an conclude in
section~\ref{sec:conclusions}.

\section{Observations and data reduction}
\label{sec:obs}

\subsection{CN and HCN}
 
The HCN $J=4-3$ and CN $N=3-2$, $J=7/2-5/2$ lines were observed towards
MWC 480 and DM Tau with the Atacama Large (sub-)Millimeter Array
(ALMA) during Cycle 0 (PI: E. Chapillon, proposal 2011.0.00629). The
Band 7 observations were carried out in October 2012 with 23
antennas. The total on-source observing time was 34~min and the
baselines lengths ranged between 21 and 384~m. The frequency bandpass
was calibrated by observing the bright quasar J0423-013 at the
begining of the observing run. The phase and amplitude temporal
variations were calibrated by regularly observing the nearby quasars
J0423-013, J01510-180 and Callisto. J01510-180 was used to derive the
absolute flux scale. Two spectral windows of 468~MHz bandwidth and
122~kHz spectral resolution were centered on the CN and HCN lines.

The data were calibrated by the ALMA staff using standard procedures,
and retrieved from the public archive. We further
self-calibrated the data in phase and amplitude in CASA, taking
advantage of the bright continuum emission in both sources. The
solutions of the self-calibration derived for the continuum were
applied to each spectral window. The continuum was then subtracted
from the visibilities using channels free of line emission to produce
the spectral line cubes. The images were obtained by deconvolution of
the visibilities using the CLEAN algorithm in CASA, with Briggs
weighting and a robust parameter of 0.5, resulting in an angular
resolution of $0.6''$. A clean mask was created for each channel to
select only regions with line emission during the cleaning
process. Table~\ref{tab:obs} summarizes the observational parameters
of the lines.

\FigRadProfiles{}

\subsection{\hthcn{} and \hcfin{}}

The \hthcn{} $J=3-2$ and \hcfin{} $J=3-2$ lines were observed towards
MWC 480 with ALMA during Cycle 2 as part of project 2013.1.00226 (PI:
K.I. \"Oberg). These observations, partially described in
\cite{oberg2015}, are part of a larger survey that did not include
DM~Tau. In brief, the Band 6 observations were carried out in June
2014 with 33 antennas with baseline lengths spanning between 18 and
650~m. The total on-source observing time was 24~min. The quasar
J0510$+$1800 was observed to calibrate the amplitude and phase
temporal variations, as well as the frequency bandpass. The same
quasar was used to derive the absolute flux scale, as no planet was
available during the observations. Two spectral windows of 59~MHz
bandwidth and 61~kHz channel spacing covered the $\hthcn$ and $\hcfin$
lines.

A description of the phase and amplitude calibration, as well as the
self-calibration, can be found in \citet{oberg2015}. The
self-calibrated visibilities were cleaned in CASA using a robust
parameter of 1.0, yielding an angular resolution of $\sim0.6''$ (see
Table~\ref{tab:obs}). This selection of robustness parameter allowed
us to improve the signal-to-noise ratio of the weak \hthcn{} and
\hcfin{} lines, while maintaining a good angular resolution. A clean
mask was created for each channel, similarly to CN and HCN, to select
only regions with line emission during the cleaning process.

\section{Observational results and analysis}
\label{sec:results}

\subsection{Spatial distribution of CN and HCN}

Figure~\ref{fig:channelmaps} shows the channel maps of the CN $N=3-2$,
$J=7/2-5/2$ and HCN $J=4-3$ lines in the MWC~480 and DM~Tau disks. The
velocity pattern characteristic of Keplerian rotation is seen for both
lines and disks. Toward MWC 480, the HCN emission is clearly more
compact than the CN emission in all channels. Such a difference in
emission patterns is not observed toward the DM Tau disk. Furthermore,
DM Tau exhibits a much more complex emission pattern in both the CN
and HCN lines. The CN `butterfly' shape in some of the central
channels is caused by spectrally resolved hyperfine components. HCN
seems to present an outer ring, whose potential origin is discussed in
section~\ref{sec:discussion}.
 
Fig.~\ref{fig:profiles} shows the continuum, and velocity integrated
maps and azimuthally averaged radial profiles of the CN and HCN lines
towards MWC~480 and DM~Tau. Toward MWC 480, these data representations
confirm the noted differences in CN and HCN emission regions in
Fig.~\ref{fig:channelmaps}. CN emission is detectable at a 2$\times$
larger radius compared to HCN. The best-fit half-light radius for the
HCN and CN emission toward MWC~480 are 66 and 102~AU, respectively. We
further note that the HCN emission has a similar size to that of the
dust continuum emission, which is shown in dashed-line in
  Fig.~\ref{fig:profiles}. In contrast, toward DM~Tau the best-fit
half-radius for HCN and CN are similar, 151 and 158~AU,
respectively. The HCN outer ring hinted at in the DM Tau channel maps
is also confirmed as a `bump' around 300~AU in the radial profile. No
such ring is seen toward MWC 480.

\FigResiduals{}
\TabParam{}
\FigChiCN{}

To quantify the HCN and CN radial abundance profiles we modeled the CN
and HCN emission using the procedure and disk density and temperature
structure described in \citep{oberg2015}. Key parameters adopted in
the model are listed in Table~2. In brief, we model
molecular abundances by defining them as power-laws
$X=X_0(r/R_0)^\alpha$, with $X_0$ the abundance with respect to total
hydrogen at $R_0=100$~AU.  We also included an outer cut-off radius,
$\Rout$. We ran grids of models for different outer radius
$50<\Rout/\mathrm{AU}<500$, power-law indexes $-2<\alpha<2$ and
molecular abundances $10^{-14}<X_{100}/\mathrm{\pscm}<10^{-9}$.  The
line emission was calculated using the radiative transfer code LIME
\citep{brinch2010} assuming non-LTE excitation with collision rates
from \citep{dumouchel2010} for HCN and from \cite{lique2010} for
CN. For the HCN isotopologues, we use the collisional rates of
HCN. The model fitting was done in the $u-v$ plane by computing
$\chi^2$, the weighted difference between model and observations, for
the real and imaginary parts of the complex visibilities. The best-fit
models were obtained by minimizing $\chi^2$.

Fig.~\ref{fig:chi_cn_hcn} shows the resulting contours of the $\Delta
\chi^2 = \chi^2-\chi^2_{min}$ distribution for the HCN and CN
abundance profiles. The best fit model of both CN and HCN lines
corresponds to a decreasing abundance profile as a function of radius,
with $\alpha=-1$. The best-fit outer-radius of the HCN emission is
110~AU, while the outer-radius of the CN emission is constrained to
$>300$~AU, \ie{} $>2\times$ larger than HCN. The best-fit abundances
at 100~AU of CN and HCN are given in Table~3 but they are highly
  model dependent, i.e. they depend strongly on the assumed vertical
abundance profile, and should not be directly compared with model
predictions.  Table \ref{tab:modelresults} lists the best fit
parameters together with what are formally 7$\sigma$ error bars based
on the data SNR. In our case the data SNR is high enough that we are
dominated by stochastic noise from the radiative transfer code and the
7$\sigma$ error bar represents the deviation from the best fit model
where there is a clear, visual misfit between model and data.

Figure~\ref{fig:residuals} shows the excellent fit between our
best-fit model and the data. Except for the low-velocity CN
channels there are no significant residuals, and those residuals are
due to the second hyperfine component of CN, which is not included in
our model. This hyperfine component is weak in MWC~480, but it is
clearly seen in DM~Tau (see Fig.~\ref{fig:channelmaps}). The good fit
is also apparent when comparing the disk-integrated spectra
(Fig.~\ref{fig:spectra_HCN}), where the spectra are extracted using a
Keplerian mask.

\FigSpectraHCN{}
\FigIsotpologues{}

\subsection{HCN isotopologues}

Figure~\ref{fig:isotopologues} displays the moment-zero maps of the
\hthcn{} and \hcfin{} lines in MWC~480 together with the integrated
emission of the blue and red shifted part of the the spectra,
illustrating the disk rotation. The emission of the two isotopologues
is compact, visually similar to the dust continuum and the HCN line
emission.  

We modeled the emission of the HCN isotopologues following the same
procedure as for CN and HCN above. Figure~\ref{fig:chi_h13cn_hc15n}
shows the $\Delta \chi^2$ distribution for the \hthcn{} and \hcfin{}
abundance profiles. The best fit parameters are listed in Table 3.
The good fit of the best-fit model to the data disk integrated spectra
(extracted using a Keplerian mask) is shown in
Fig.~\ref{fig:spectra_Iso}. The best-fit model abundances at 100~AU
are given in Table~3. Similarly to CN and HCN, the absolute
abundances are uncertain, due to an unknown vertical abundance
profile. Assuming that the HCN isotopologue emission originates in the
same disk layer and that the lines are optically thin, their abundance
ratio should be robust, however. This abundance ratio is 2.8$\pm$1.4
and holds for the entire disk out to 100~AU since both lines are best
fit by the same power-law index of $-0.5$.

\FigSpectraIso{}

In the MWC 480 disk, the integrated $\hthcn/$HCN density flux ratio is
$\sim36$, which is lower than the $^{12}$C/$^{13}$C isotopic ratios
observed in the ISM, indicating that the HCN line is partially
optically thick. We therefore use \hthcn{} as a proxy of HCN to
calculate the nitrogen fractionation in HCN. Assuming an isotopic
ratio of $^{12}$C/$^{13}$C$=70$, we derive a disk HCN isotopologue
ratio HC$^{14}$N/$\hcfin=200\pm110$. To estimate this error we
included a 30\% uncertainty in the C isotopic ratio, as this value has
been shown to vary both with time and density \citep{roueff2015}. The
main uncertainty ($\sim85$\%), however, arises from the error in the
measured $\hthcn/\hcfin$ ratio.

\FigChiISotop{}
\TabModelResults{}

\section{Discussion}
\label{sec:discussion}

\subsection{Photochemistry: CN and HCN}

The observed CN and HCN disk emission profiles in the MWC 480 disk are
well-fit by power-law abundance profiles with a radial cut-off that is
two times larger for CN compared to HCN. CN is thus significantly
more radially extended than HCN, \ie{} we can exclude with a high
level of confidence that excitation alone is responsible for the
different HCN and CN emission profiles. This result is consistent with
chemical model predictions of survival of CN and rapid
photodissociation of HCN in the outer, low-density disk
\citep{aikawa2002,jonkheid2007}.  Indeed, these models predict that
the CN abundance can remain constant and even increase with radius
because radical formation is favored at low densities, and because CN
is a major HCN photodissociation product. HCN, on the other hand,
should only be present in the high-density parts of the disk where a
shielded molecular layer exists.

The same models also predict a different vertical structure of CN and
HCN, with the CN abundance peaking close to the disk surface in the
inner disk regions. Constraining the vertical abundances of these
molecules requires multiple line transitions and/or direct
observations of vertical emission patterns in edge-on disks. In the
meantime, the derived abundance ratio of the two molecules remains
relatively unconstrained.

The emission of CN and HCN in the disk around T tauri star DM Tau
exhibit different patterns compared to MWC~480. First, the spatial
distribution (the extension of the emission) is similar for CN and HCN
in the DM~Tau disk.  This difference between MWC 480 and DM~Tau may be
explained by the fact that the UV field around DM~Tau should be
substantially lower. Less vertical disk column may therefore be
required to shield the outer disk regions, resulting in a shielded
molecular layer out to many 100s of AU in the DM~Tau disk.  Second,
HCN exhibits a low SNR outer ring around 300~AU, best identified in
the azimuthally-averaged emission profile. The gap demarcating the
main HCN disk from the outer ring is faintly seen in CN as well. This
gap may be the result of micron-sized dust depletion at the same
radius due to dust dynamics, which would enable UV radiation to
penetrate down toward the midplane, photo-dissociating HCN. Such a
dust lane has been observed toward TW Hya in scattered light
\citep{debes2013}. Similar high-quality scattered light observations
toward DM Tau could resolve this question. If this scenario is
confirmed, these ALMA observations suggest that CN/HCN observations
could be developed into a powerful probe of dust distributions and
dynamics in the outer regions of disks.

\subsection{Nitrogen fractionation in the MWC480 disk}   

\FigIsotopologuesRatios{}

We used observations of HCN isotopologues to determine the level of
nitrogen fractionation in the MWC 480 disk. Fig.~\ref{fig:ratios}
shows the measured \Nratio{} ratio in the MWC 480 disk compared with
different Solar system bodies and the ISM. Observations of the
\Nratio{} ratio in our Solar System reveal a clear difference between
gaseous and rocky bodies \citep{mumma2011}. A low N fractionation
(\ie{}, high \Nratio{}) is observed toward the Sun and Jupiter
\citep{marty2011}, while a high fractionation (\ie{}, low \Nratio{})
is observed in the rocky planets, comets and meteorites
\citep{bockelee2008}. Isotopic ratios in the coma of comets are
determined remotely through high-resolution radio and optical
spectroscopy. The ROSETTA space mission will hopefully provide {\it
  in-situ} measurements of the $\Nratio$ ratio in the coma of comet
67P/Churyumov-Gerasimenko. Comets display the highest $^{15}$N
enhancements among Solar System bodies \citep{mumma2011}. Prior to
this study the only other place where such high nitrogen fractionation
is found is cold pre-stellar cores
\citep{hily-blant2013,wampfler2014}. The cometary $^{15}$N abundance
could thus signify an interstellar inheritance.

Interstellar inheritance is not the only possibility,
however. Distinguishing between pre-Solar and Solar Nebula origins of
nitrogen fractionation requires observations of nitrogen fractionation
in protoplanetary disks.  In general, HCN (isotopologue) formation
begins with atomic nitrogen.  At low temperatures $^{15}$N is
preferentially incorporated into cyanides resulting in
fractionation. Observations of HCN and HNC fractionation toward
protostars present a tentative trend of the \Nratio{} with
temperature, supporting this scenario \citep{wampfler2014}. The second
mechanism to produce an enhancement in $^{15}$N in HCN is isotope
selective photodissociation of $^{14}$N$^{15}$N over $^{14}$N$_2$,
resulting in an enhanced atomic $^{15}$N/$^{14}$N ratio. Theoretical
models have shown that this mechanism can increase the \Nratio{} ratio
by up to a factor 10 in protoplanetary disks \citep{heays2014}.

Pre-stellar inheritance, low-temperature chemical fractionation in
cold disk regions, and photochemically driven fractionation in disk
atmospheres could thus all explain the observed nitrogen fractionation
in comets. The presented HCN isotopologue observations in the MWC~480
disk cannot distinguish between these scenarios. Future, higher SNR
and/or higher resolution disk studies could, however.  The three
scenarios should present very different radial and vertical \Nratio{}
profiles. Inheritance should produce an almost uniform ratio across
the disk. Low-temperature chemical fractionation should result in a
lower \Nratio{} ratio in HCN in the outer, cold disk that is still
protected from radiation and also a decreasing ratio toward the disk
midplane. Finally, nitrogen fractionation from selective
photodissociation should result in the lowest \Nratio{} ratio in HCN
closer to the disk atmosphere. In addition to detailed studies of
individual sources, observations of the averaged \Nratio{} ratio in
disks exposed to different stellar radiation fields will also bring
insight into the origin of N fractionation. If low-temperature
chemical fractination is the dominant mechanism producing \hcfin{},
then a multi-source study should result in a trend of increasing
\Nratio{} ratio with stellar radiation field. In contrast, the
opposite behaviour should be observed if most of the \hcfin{} is
produced by selective photodissociation in the surface layers. These
observations will reveal how frequently protoplanetary disks are
seeded with $^{15}$N enriched molecules, and whether additional
enrichment occurs in the disk phase, changing the fractionation
pattern during planet formation. The clear HC$^{15}$N detection in MWC
480 after 20 min integrated demonstrates that such observations will
be readily achievable toward a sample of disks in a reasonable time
frame.

\section{Summary and Conclusions}
\label{sec:conclusions}

Spatially and spectrally resolved ALMA observations toward the
protoplanetary disk surrounding Herbig Ae star MWC~480 clearly show
that its CN emission is $\sim2$ times more extended than its HCN
emission. We constructed a parametric model of the disk to obtain the
underlying abundance profile of the two species and found that the
observed difference in emission profiles correspond to similarly
different abundance profiles for CN and HCN. This confirms
theoretical model predictions which show that HCN is only abundant in
disk regions where the radiation field has been attenuated
sufficiently to prevent its photodissociation, while CN can survive in
the outer low-density, unshielded disk regions. This difference in CN
and HCN radial extent is not seen toward the disk around T Tauri star
DM Tau, suggestive of different photochemical structures in disks
around low and intermediate mass stars. The DM Tau disk also presents
an intriguing outer ring in HCN (and potentially in CN). Observation
of CN and HCN coupled to detailed modeling clearly have great
potential as tracers of UV radiation field strengths across
protoplanetary disks.

We also presented the first detection of the \hthcn{} and \hcfin{}
isotopologues in a protoplanetary disk. We modeled the \hthcn{} and
\hcfin{} emission, and derive a disk average for the \Nratio{}
isotopic ratio of $200\pm100$ in MWC~480. This value, which is lower
than the Solar one, is similar to what has been measured in comets and
in the cold ISM. These observations show that with the current ALMA
capabilities it is remarkable easy to detect the faint lines of rare
istopologues of common molecules in protoplanetary disks making new
detections in the near future very likely. Observations at higher
sensitivity in MWC~480, as well as in other sources, should allow us
to resolve variations of the \Nratio{} ratio across the disk, and
directly address the current fractionation puzzle of why different
objects in our Solar system present different N enrichment levels.

\end{document}